# Stokes Vector Spectroscopy of Nonlinear Depolarized Light

*Lothar Moeller*

*SubCom, Eatontown, NJ 07716, USA*
*lmoeller@subcom.com*

**Abstract**

Experimental Stokes vector spectroscopy characterizes polarization state noise in nonlinear fiber transmission. Noise bandwidths of a few MHz are observed from XPM-blurred receive SOPs using a polarization scrambling interferometer. This method can detect data pattern correlations among WDM channels that contribute to performance degradations in signalling.

## 1   Introduction

In general, nonlinear (NL) transmission phenomena are polarization-dependent. NL polarization rotation [1], utilized in Kerr shutters [2,3], stands out among them, as its elegant formalism in Stokes space explicitly reveals the dependence of a signal's state of polarization (SOP) on the fiber nonlinearity. Recently, a novel polarization-dependent NL transmission phenomenon which we refer to as NonLinear DePolarization (NLDP) of light in optical fiber has been observed [4]. While co-propagating, unpolarized broadband light such as ASE depolarizes a cw light, which is fully polarized at its launch. In contrast to NL polarization rotation where both involved signals possess at least quasi deterministic SOPs, NLDP is an interaction between a probe with deterministic SOP and a signal with completely random polarization (i.e. ideally 0 DOP) bearing significant consequences.

While NLDP causes small transmission penalties, we see its main relevance as being a vehicle for testing limits of polarization-dependent NL propagation theories. For example, the NL coupled Schrodinger equations (NCSE) [5] accurately describe NL signal interactions within a linear birefringent waveplate. And the Manakov-PMD equation [6], a formal approach to combine the NL features of a fiber and its birefringence character, is widely believed to cover all propagation aspects of SSMFs including polarization effects but is usually difficult to solve. Its simplified version, known as simply the Manakov equation [7], was originally introduced heuristically and later formally derived by incorporating an effective NL perturbation term into the NCSE [5]. Currently this approach is often applied in system modelling. While it can properly describe phenomena such as NL polarization rotation its application does not lead to a correct 1$^{st}$ order approximation for NLDP.

We have demonstrated NLDP in a commercial undersea communication link and lab setups based on recirculating fiber loops (RFL) by recording the histograms of SOP speed of a transmitted probe [4]. Here, we report for the first time the power spectral density of the probe's Stokes vector when undergoing NLDP, using a novel polarization scrambling interferometer (PSI) technique. This enhances physical understanding of the process' temporal and spectral features which is not only useful for aforementioned testing of propagation theories but required for future signal equalization techniques.

## 2   Homodyne polarization interferometry of NLDP

NLDP appears as small but ultra-fast SOP fluctuations with speeds in the 10's of Mrad/s for typical experimental conditions. A weak cw probe launched at the link's input experiences both, SOP fluctuations and phase jitter that result from NL interactions with unpolarized neighbouring signals. At the link output, the probe's field with constant power is described by a time-dependent normalized Stokes vector $\vec{s}$ and a polarization-independent phase $e^{j\varphi(t)-i\omega_0 t}$, where $\omega_0$ is the angular optical frequency of the cw light. Since NLDP occurs as small SOP fluctuation, we rewrite the probe's Stokes vector $\vec{S_0} + \vec{\sigma}(t)$ as the sum of a slowly varying Stokes vector $\vec{S_0}$ and a short but fast-changing vector $\vec{\sigma}(t)$.

After transmission, the probe enters a polarization scrambling interferometer (PSI, Fig. 1) with two paths ('A' and 'B') containing an adjustable delay line to match their optical lengths to within a few millimeters. An acousto-optical modulator (AOM), operated in cw mode, shifts the frequency of a traversing signal by 27.1 MHz; and thus after combining the two equally strong signals of both arms (VOA, variable optical attenuator) a beat tone with sufficient frequency spacing relative to DC can be detected by a RF spectrum analyser (ESA). A polarization scrambler in one arm slowly rotates (~10 rad/s) the SOP resulting in interferences of both path outputs at random polarization superpositions, which benefits a mean value analysis. To good approximation, the AC photocurrent i(t) of the beat signal and its autocorrelation $\varphi_{ii}(\tau)$ in time domain, which yields the power density spectrum of the probe's Stokes vector, can be expressed as:

$$i(t) \propto \sqrt{1 + \vec{S}_A(t)\vec{S}_B(t)}\ e^{-j\omega_{AOM}t}e^{j\vartheta};$$

$$\varphi_{ii}(\tau) = \overline{i(t)i(t+\tau)^*} \propto (\frac{1}{2} + \frac{1}{3}\overline{\vec{\sigma}(t)\vec{\sigma}(t+\tau)})\cos(\omega_{AOM}\tau) \quad (1)$$

where t, $\vartheta$, and $\overline{\phantom{x}}$ stand for the time, a phase, determined by experimental conditions, and time averaging, respectively. The ACF in 1st o approximation follows after averaging over the Poincare sphere and holds for $1 + \overline{S_{A0}}\ \overline{S_{B0}} \gg \overline{\vec{\sigma}(t)\vec{\sigma}(t)}$. Since $\vec{\sigma}_A(t)$ and $\vec{\sigma}_B(t)$ are relatively short vectors, most surface parts of the Poincare sphere fulfill this condition except for a small sold angle where $\overline{S_{B0}}$ approaches $-\overline{S_{A0}}$ as exemplarily shown in Fig.1 for a Stokes vector $\overline{S'_{B0}}$.

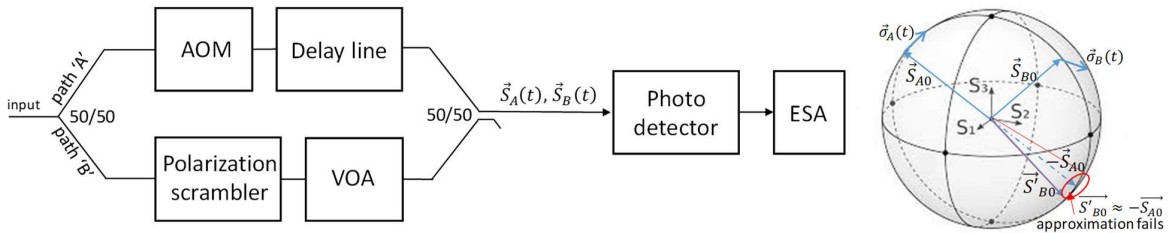

Fig. 1: Polarization scrambling interferometer for Stokes vector spectroscopy. Averaging over all possible $\vec{S_{A0}} + \vec{\sigma}_A(t)$, $\vec{S_{B0}} + \vec{\sigma}_B(t)$ arrangements yields the Stokes vector spectrum. Derivation holds if $\vec{S_{A0}}$, $\vec{S_{B0}}$ diverge from quasi anti-parallel settings and reside outside the shown small red circle.

## 3 Experimental setup and results

Except for the PSI, which substitutes a polarimeter in [4], we use the same test bed (Fig. 2a) and experimental conditions as described in Section III A of [4]. A probe (generated by an external cavity laser, ECL) is combined with a spectrally flat ASE loading (Fig.2b) and propagates over adjustable transmission distances after entering the RFL via path I. Its Stokes vector spectrum (Fig. 2c) is recorded using a PSI which contains a gated RF spectrum analyser (ESA) with 30 kHz resolution. In order to demonstrate that NL transmission effects widen the pedestal of the Stokes vector spectrum, we follow the same comparison strategy as explained in [4] and launch a reference signal via path II (btb measurement) while detecting it at equal OSNR and optical power (Fig. 2d) as the probe possesses. Here, no broadening of the pedestal appears, and signal-ASE beat noise determines the noise floor. Transmission distance-dependent gain ripples of the RFL lead to slightly different received optical probe powers. For an easier comparison, we plot the RF spectra with normalized peak powers (0 dB). To analyze the spectra, we subtract corresponding btb measurements from the transmitted spectral density and perform a Lorentzian curve fitting. The derived FWHMs, shown in Fig.2c, shrink with increasing transmission distance, confirming theoretical predictions (Eqn7 and 8 [4]). The NLDP contributions of each span add partially coherently and sharpen the Stokes vector spectrum as a function of distance. The auto-correlation (Eq.1) is derived by assuming that all vector cross-correlations cancel with sufficient averaging. Therefore, each RF spectrum consists of ~8000 averaged scans recorded over a measurement period of ~1.5 h. To rule out the possibility that the spectral pedestals in Fig. 2 c) are caused by amplitude modulation (AM) (e.g. FWM), we launch an AM probe into the RFL. This probe is generated by sinusoidally modulating the ECL output with a LiNOb$_3$ MZ modulator (biased at zero point and operated at half of the AOM frequency). Path A of the PSI is blocked and Fig. 3e) shows the recorded RF spectra of the probe after transmission and for the corresponding reference. No significant broadening of the pedestal is visible, thus any AM of the probe undergoing NLDP is small. The spikes (~50 dB suppressed), in both spectra stem from control loop electronics that stabilize the drifting LiNOb$_3$ modulator. The optical probe powers were matched on the receive side yielding higher SNR in the RF spectrum for the AM probe.

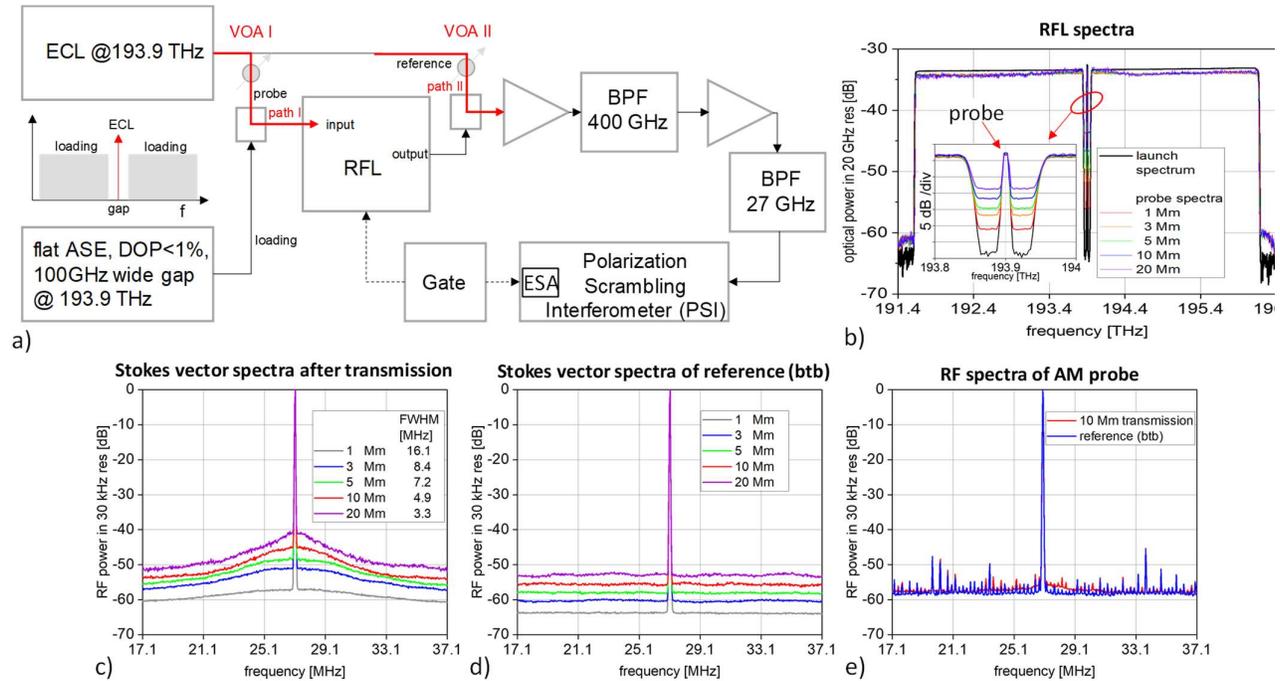

Fig. 2: a) Recirculating fiber loop (RFL) test bed with a Polarization Scrambling Interferometer (PSI). A probe (ECL), embedded by unpolarized flat ASE, propagates through the RFL and the PSI records the power density spectrum of its Stokes vector. b) Typical launch and receive optical spectra. c) Stokes vector spectra for 1 to 20 Mm transmissions. d) Stokes vector spectra of reference (btb) show flat noise floor. e) Similar reference and transmit RF spectra of AM probe indicate negligible FWM.

## 4 Nonlinear depolarization from (un)-correlated unpolarized WDM signals

The impact of correlated dummy channels, emulating traffic load in lab experiments, has been discussed in terms of both simulations and experiments by analysing BER degradations of received data signals [8]. NLDP blurs a signal's receive SOP which contributes to the performance degradation via coherent crosstalk. Here, we compare the depolarization caused by three different dummy loading schemes using Stokes vector spectroscopy. Dummy loadings I and II consist of 60 x 200 Gb/s PM-DQPSK signals at 56 GBd with a frequency spacing of 62.5 GHz. In the highly correlated load I, every second WDM channel carries the same data at launch.

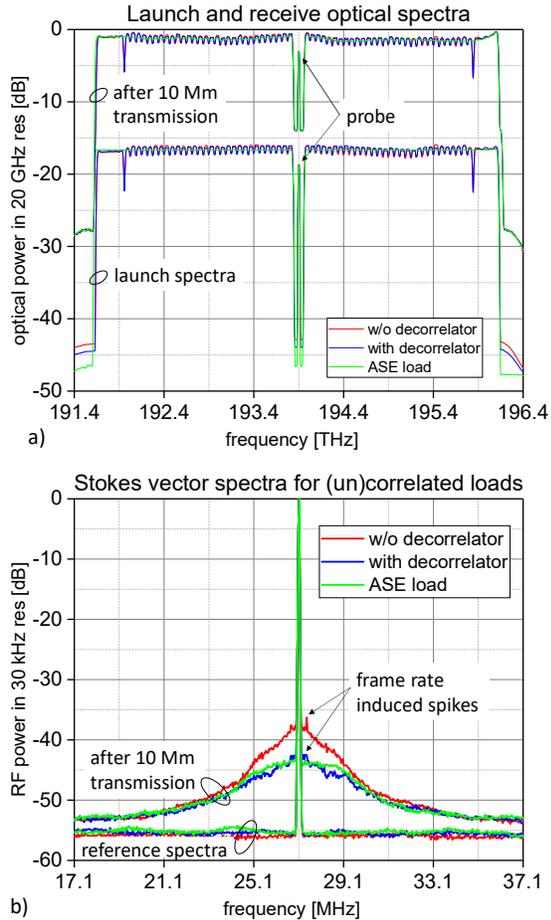

The less correlated load II is obtained by demuxing load I, introducing individual channel delays (several 10s of m fibers), and multiplexing thereafter all channels together (de-correlator). Both loads possess a 125 GHz wide gap, centered at 193.9 THz to accommodate the probe, and ASE-packed edges to fill up the C-band. The uncorrelated load III is formed by unpolarized ASE with identical edge and gap characteristics to loads I and II. All flat load spectra at launch (limited by technical realization constraints) maintain their shape after 10 Mm transmission due to the test bed's flat transfer function (Fig. 3a). The Stokes vector spectrum, generated by load I (highly correlated) possesses the highest RF power, indicating stronger NLDP than the two other loading schemes are causing (Fig. 3b). Loads II and III generate Stokes vector spectra with similar RF power indicating a high degree of de-correlation of load II. Interestingly, close to the carrier, frequency spikes at $27.1 \pm 0.2$ MHz are visible for load I and II that correlate with the frame repetition rate of the data signals (200 kHz). By comparing the signatures of loads I and II, we conclude that correlated data channels generate perturbations that partially add coherently and yield stronger NLDP. Thus, our method has the potential of identifying degrees of correlation among WDM loads and can be useful for analysing the origin of XPM-induced transmission penalties.

## 5 Conclusion

By means of a novel polarization scrambling interferometer we performed Stokes vector spectroscopy of probe signals undergoing NL transmission in long haul systems. Self-interference of the probe under random SOPs and upon statistical averaging over the Poincare sphere yields the Stokes vector power spectral density that characterizes polarization state noise. With this interferometric approach we measured the Stokes vector spectrum of a nonlinear depolarized probe and found noise bandwidths in the few MHz range after transpacific transmission distances which agrees with previously published theory [4]. The Stokes vector spectra indicate correlations among dummy channels, commonly used for traffic load emulations in lab experiments, and can be useful for studying origins of transmission penalties.

## 6 References

[1] L.F. Mollenauer et al., Opt. Lett. p. 2060, 1995.
[2] K. Kitayama et al., Appl. Phys. Let. 46, p. 623, 1985.
[3] L. Moeller et al., PDP20, OFC 2004.
[4] L. Moeller, arXiv:1902.05168, 2019.
[5] G. Agrawal, Nonlinear Fiber Optics, 3rd ed., Academic Press, p. 207, 2001.
[6] D. Marcuse et al., J. Lightwave Technol., p. 1753, 1997.
[7] S.V. Manakov, Sov. Phys.-JETP 38, p. 248, 1974.
[8] R. Dar et al., ECOC'16, p. 483, 2016.

Fig. 3: a) Flat launch and receive optical spectra of an ASE load and (un)-correlated data channels emulating traffic (offset added for better visibility). b) Stokes vector power density spectra of a probe widen by NL interaction with aforementioned loads. Correlation among dummy channels increases the polarization noise of a received probe after 10 Mm transmission.